\documentclass[aps,prl,showpacs,preprint]{revtex4}
\usepackage{times}
\usepackage[T1]{fontenc}
\usepackage[latin1]{inputenc}
\usepackage{amsmath}
\usepackage{graphics}
\usepackage{subfigure}

\begin{document}

\pacs{75.10.Hk, 75.30.Cr, 75.40.Cx}

\title{Classical generalized constant coupling model for geometrically frustrated
antiferromagnets}

\author{Angel J. García-Adeva}

\email{garcia@landau.physics.wisc.edu}

\author{David L. Huber}

\affiliation{Department of Physics; University of Wisconsin--Madison; Madison, WI 53706}

\begin{abstract}
A generalized constant coupling approximation for classical geometrically frustrated
antiferromagnets is presented. Starting from a frustrated unit we introduce
the interactions with the surrounding units in terms of an internal effective
field which is fixed by a self consistency condition. Results for the magnetic
susceptibility and specific heat are compared with Monte Carlo data for the
classical Heisenberg model for the pyrochlore and \emph{kagomé} lattices. The
predictions for the susceptibility are found to be essentially exact, and the
corresponding predictions for the specific heat are found to be in very good
agreement with the Monte Carlo results.
\end{abstract}
\maketitle
\begin{figure}
{\par\centering \subfigure[\label{fig.pyro}Pyrochlore lattice.]{\includegraphics{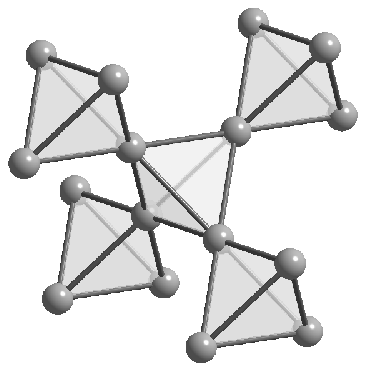}} \subfigure[\label{fig.kagome}\emph{Kagom\'{e}} lattice.]{\includegraphics{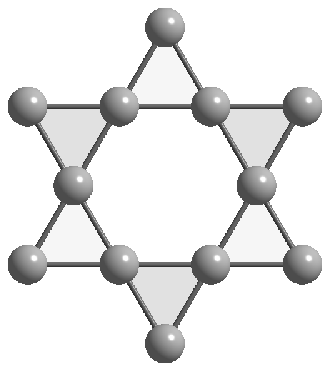}} \par}
\caption{\label{fig.structures}The magnetic lattices considered in this work.}
\end{figure}

\emph{Introduction.-} In the last several years, geometrically frustrated antiferromagnets
(GFAF) have emerged as a new class of magnetic materials with uncommon physical
properties and have received a vast amount of attention (see \cite{hfm.2000}
and references therein). In these materials, the elementary unit of the magnetic
structure is the triangle, which makes it impossible to satisfy all the antiferromagnetic
bonds at the same time, with the result of a macroscopically degenerated ground
state. Examples of GFAF are the pyrochlore and the \emph{kagomé} lattices. In
the former, the magnetic ions occupy the corners of a 3D arrangement of corner
sharing tetrahedra; in the later, the magnetic ions occupy the corners of a
2D arrangement of corner sharing triangles (see Fig. \ref{fig.structures}).
In the case of materials which crystallize in the pyrochlore structure, the
magnetic susceptibility follows the Curie--Weiss law down to temperatures well
below the Curie--Weiss temperature. At this point, usually of two orders of
magnitude smaller than the Neél point predicted by the standard mean field (MF)
theory, some systems exhibit some kind of long range order (LRO), whereas others
experience a transition to a spin glass state (SG). This is a striking feature
for a system with only a marginal amount of disorder. Finally, there some pyrochlores
which do not exhibit any form of order whatsoever, and are usually regarded
as spin liquids. In the case of the \emph{kagomé} lattice, even though there
are very few real systems where this structure is realized, the magnetic properties
fall in two great categories: the vast majority of the compounds studied show
a transition to a LRO state with non collinear configuration of spins, and a
few systems exhibit no LRO, but a SG like transition.

For these reasons, it is easy to understand the large amount of attention, both
from the experimental and the theoretical point of view, these systems have
attracted. From the theoretical point of view, a number of techniques have been
used to try to understand the origin of the puzzling properties mentioned above.
All the theoretical results seem to indicate that for these geometries the classical
Heisenberg model with only nearest neighbor interactions does not order at any
finite temperature, in agreement with Monte Carlo (MC) results \cite{moessner.1999,moessner.1998,Reimers.1992,Chalker.1992,Reimers.1993,Garanin.1999}.
There are also a relatively few works which have dealt with the quantum effects
in these systems \cite{harris.1991}. However, the main interest during last
years has been in the classical GFAF.

In any case, the current theories have to deal with two main problems: the first
one is that in real systems there are additional effects, which can be dipole--dipole
interactions, small anisotropies, or further neighbor interactions, which are
not easy to include in the theory, and so make it very difficult to compare
with experimental results. However, this fact can be circumvented by using MC
data calculated for the particular assumptions of the model. The second problem
is that these models only provide qualitative agreement even when compared with
MC data. In this sense, the single unit approximation, is the model which probably
gives the best quantitative agreement when compared with susceptibility MC data
for the pyrochlore lattice \cite{moessner.1999}. However, in order to reach
such a good agreement, it is necessary to introduce an \emph{ad hoc} rescaling
procedure with an unclear physical justification. In the other hand, the infinite
component spin vector model provides qualitative agreement for the \emph{kagomé}
case \cite{Garanin.1999}.

Obviously, the difficulty to find reliable models is rooted in the complexity
of the geometry of these systems. Therefore, we think that, before embarking
on the development of complicated models (which receive small support even from
MC data obtained for the particular assumptions of the model), we should try
to find simpler reliable approximation schemes which, as a first step, are in
good qualitative and quantitative agreement with MC data. At the same time,
these models should be as simple as possible (mathematically speaking), in order
to be easily generalized to include additional interactions that play an important
role in real systems. Only in this way, can we compare the models with data
obtained from real systems, without this comparison being obscured by the complexity
of the mathematical formalism.

In this work, we deal with the first of the two steps mentioned in the previous
paragraph. We introduce a generalization of the well known constant coupling
(CC) method \cite{kastel.1956}, which has been successfully applied to the
study of 3D standard ferro and antiferromagnets. Our method, which we will call
generalized constant coupling (GCC) method, takes into account the geometrical
peculiarities of the frustrated structures mentioned above. In spite of its
mathematical simplicity, the susceptibilities and specific heat calculated for the pyrochlore
and \emph{kagomé} lattices, in the paramagnetic region, are in excellent agreement
with MC data for the classical Heisenberg model in these systems. 

It is important to note that, even though we have focused the discussion of the present work in the context of geometrical frustration in Heisenberg systems, this approach represents, in fact, a general technique for the investigation of the thermodynamic properties of spin Hamiltonians in frustrated geometries and, thus, its applicability is not limited to the GFAF, but could be useful in the investigation of other physical systems where geometrical frustration is relevant (see ref.\ \onlinecite{hfm.2000} for a discussion of other problems where geometrical frustration is relevant).

\emph{The model.-} The Heisenberg model with only nearest neighbor (NN) interactions
in the presence of a magnetic applied field \( H_{0} \) is described by the
Hamiltonian \cite{smart.1966}
\begin{equation}
H=J\sum _{\left\langle i,j\right\rangle }\vec{s}_{i}\cdot \vec{s}_{j}-H_{0}\sum _{i}s_{z_{i}},
\end{equation}
 where \( J \) is the positive antiferromagnetic coupling, \( \vec{s}_{i} \)
and \( \vec{s}_{j} \) represent classical spins of modulus \( s_{0} \) located
in a pyrochlore or \emph{kagomé} lattice and \( s_{z_{i}} \) the corresponding
component along the applied field, and the sum is done over pairs of NN.

The idea of our approximate method is based on the experimental fact that the
spin-spin correlations of the classical Heisenberg in the pyrochlore lattice
are always short ranged \cite{moessner.1999}. Therefore, it is a reasonable
approximation to start by considering isolated units (tetrahedra or triangles,
for the pyrochlore and \emph{kagomé} lattices, respectively), and later add
the interactions with the surrounding units by in an approximate way (this is
in contrast with the standard CC method, in which the elementary magnetic unit
is taken as a pair of magnetic ions). Thus, it is important to first study the
properties of the individual units. This task has been carried out by Moessner
and Berlinsky \cite{moessner.1999}, and the result obtained for the partition
function of an isolated unit with \( p \) spins in the absence of applied field
is given by
\begin{equation}
Z_{p}=\frac{(4\pi )^{p+1}}{(4\pi \beta \widetilde{J})^{3/2}}\int _{0}^{\infty }x^{2}\left( \frac{\sin x}{x}\right) ^{p}\exp \left( -\frac{x^{2}}{2\beta \widetilde{J}}\right) \, dx,
\end{equation}
 where \( \beta =1/T \) (we will use units of the Boltzmann constant \( k_{B} \)
throughout this work, so the energies are expressed in absolute temperature
units), and we have introduced the effective coupling \( \widetilde{J}=Js_{0}^{2} \).
Expressions for units with 2, 3, and 4 ions can be found in the original work
by those authors.

The susceptibility per spin, \( \chi _{p} \), can be easily calculated from
the partition function by using the fluctuation--dissipation theorem, and is
given by
\begin{equation}
\label{suscep.per.spin}
\chi _{p}(\widetilde{T})=\frac{\left\langle S_{p}^{2}\right\rangle }{3pT}=\frac{2\widetilde{T}}{3pJ}\frac{\partial }{\partial \widetilde{T}}\ln Z_{p},
\end{equation}
 where \( \left\langle S_{p}^{2}\right\rangle  \) represents the thermal average
of the total spin of the unit with \( p \) ions, and we have introduced the
adimensional temperature \( \widetilde{T}=\frac{T}{\widetilde{J}} \).

Next, we introduce the interaction with neighboring units as an unknown internal
effective field, \( H_{1} \), created by the \( (p-1) \) NN ions outside the
unit. The CC approximation consists of taking this internal field as \( H_{1}=(p-1)H' \),
where \( H' \) is the average internal field acting on a spin due to each of
its NN. For example, in the case of the standard CW model, the internal effective
field is given by \( H_{1}=2(p-1)H' \), as the ions are considered separately,
and each has \( 2(p-1) \) NN in the corner sharing structures considered in
this work (see fig. \ref{fig.structures}). The internal field is evaluated by
imposing the self consistent condition of equating the magnetization per spin
in the field with that of a unit in the field, which can be mathematically stated
as
\begin{equation}
\label{consistency.condition}
s_{0}L(s_{0}(H_{0}+2(p-1)H')/T)=\frac{m_{p}(H_{0}+(p-1)H')}{p},
\end{equation}
where the left side of the equation corresponds to the value of the magnetization
per spin in the classical limit of the Curie--Weiss model, with \( L(x)=\coth x-\frac{1}{x} \)
the so called Langevin function \cite{smart.1966}. The right side corresponds
to the magnetization per spin of the isolated unit under the influence of the
internal field. In the general case, equation \eqref{consistency.condition}
can only be solved numerically. However, in the paramagnetic regime \cite{observation},
we can put, for small fields, \( \left( H_{0}+H_{1}\right) /T\ll 1 \),
\begin{equation}
\label{first}
\frac{m_{p}(H_{0}+(p-1)H')}{p}\approx \chi _{p}(T)\, (H_{0}+(p-1)H'),
\end{equation}
 for the magnetization per spin in the cluster, where the susceptibility per
spin is given by expression \eqref{suscep.per.spin}, and
\begin{equation}
\label{second}
s_{0}L(s_{0}(H_{0}+2(p-1)H')/T)\approx \frac{s_{0}^{2}}{3T}(H_{0}+2(p-1)H').
\end{equation}
 Taking into account eqs. \eqref{first} and \eqref{second}, we can solve equation
\eqref{consistency.condition} very easily in terms of the function
\begin{equation}
\varepsilon _{p}(\widetilde{T})=\frac{\left\langle S_{q}^{2}\right\rangle }{p\, s_{0}^{2}}-1=\frac{2\widetilde{T}^{2}}{p}\frac{\partial }{\partial \widetilde{T}}\ln Z_{q}(\widetilde{T})-1,
\end{equation}
 to give \( H'=\frac{\varepsilon _{p}(\widetilde{T})}{(p-1)\left[ 1-\varepsilon _{p}(\widetilde{T})\right] }H_{0} \),
and the corresponding susceptibility per ion is then given by
\begin{equation}
\chi ^{gcc}_{p}(\widetilde{T})=\frac{1}{3\widetilde{T}}\frac{1+\varepsilon _{p}(\widetilde{T})}{1-\varepsilon _{p}(\widetilde{T})}.
\end{equation}

Another interesting quantity we can readily evaluate in this model is the specific
heat in zero applied field. In this limit, the internal field will be identically
zero, and the specific heat will be identical to the specific heat calculated
for non interacting units. Taking into account the relation \( \left\langle \vec{s}_{i}\cdot \vec{s}_{j}\right\rangle =\frac{s_{0}^{2}}{p-1}\varepsilon _{p}(\widetilde{T}) \)
it can be easily shown that the specific heat per spin is given by
\begin{equation}
C_{p}=\frac{\partial }{\partial \widetilde{T}}\varepsilon _{p}(\widetilde{T}).
\end{equation}

\begin{figure}
{\par\centering \subfigure[\label{fig.suscep.pyro}Susceptibility of the pyrochlore lattice. The inset shows a detailed comparison at very low temperatures.]{\includegraphics{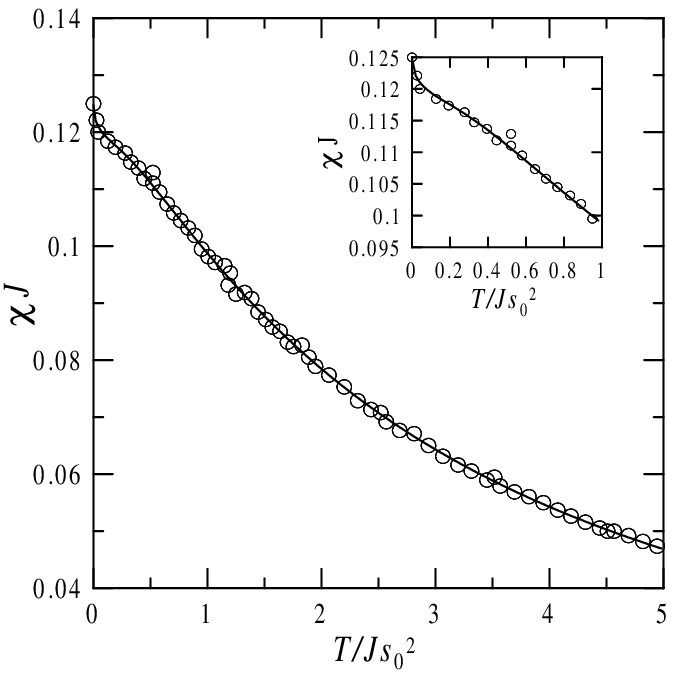}} \subfigure[\label{fig.suscep.kagome}Susceptibility of the \emph{Kagom\'{e}} lattice.]{\includegraphics{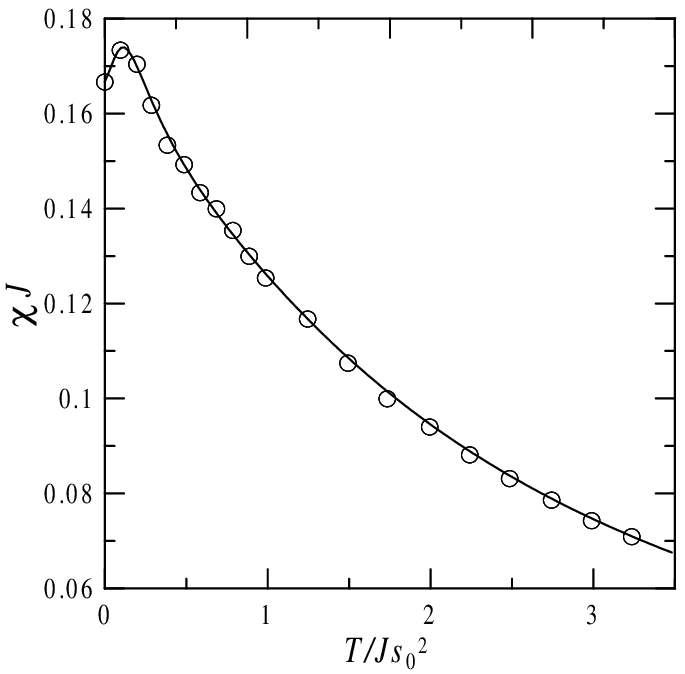}} \par}
\caption{Susceptibility from Monte Carlo calculations and the one predicted by the GCC
model. Open circles are from MC simulations (obtained from \cite{moessner.1999}
for the pyrochlore lattice and from \cite{Reimers.1993} for the \emph{kagomé}
lattice). The solid line corresponds to the GCC model.}
\end{figure}

\emph{Comparison with MC data.-} Let us first consider the susceptibility obtained
from the classical GCC model for the pyrochlore lattice. As can be seen from
the observation of Fig. \ref{fig.suscep.pyro}, the agreement between the theoretical
curve and MC data \cite{moessner.1999} can be considered as exact in whole
the temperature range. Moreover, at \( T=0 \) K, the value predicted by our
model is \( 1/8\widetilde{J} \), which is the value obtained from MC calculations.
In the inset of that figure, we can see how the theory even predicts the kink
in the susceptibility as one approaches very small temperatures.

In oder to corroborate that this surprising agreement is not a coincidence,
we also compared the susceptibility calculated from the GCC model with MC data
for the \emph{kagomé} lattice \cite{Reimers.1993}, which can be seen in Fig.
\ref{fig.suscep.kagome}. Again, the curve predicted by our model can be considered
as exact in whole the temperature range. Moreover, the value predicted by this
model at \( T=0 \) K is \( 1/6\widetilde{J} \), again the exact value.

It is important to stress that the curves depicted in Figs. \ref{fig.suscep.pyro}
and \ref{fig.suscep.kagome} \emph{have not been rescaled in any sense, nor contain
any fit parameter}. As far as we know, no other model is able to predict a such
accurate susceptibility in the whole temperature range, considering both the
pyrochlore and the \emph{kagomé} lattice on an equal footing.

\begin{figure}
{\par\centering \includegraphics{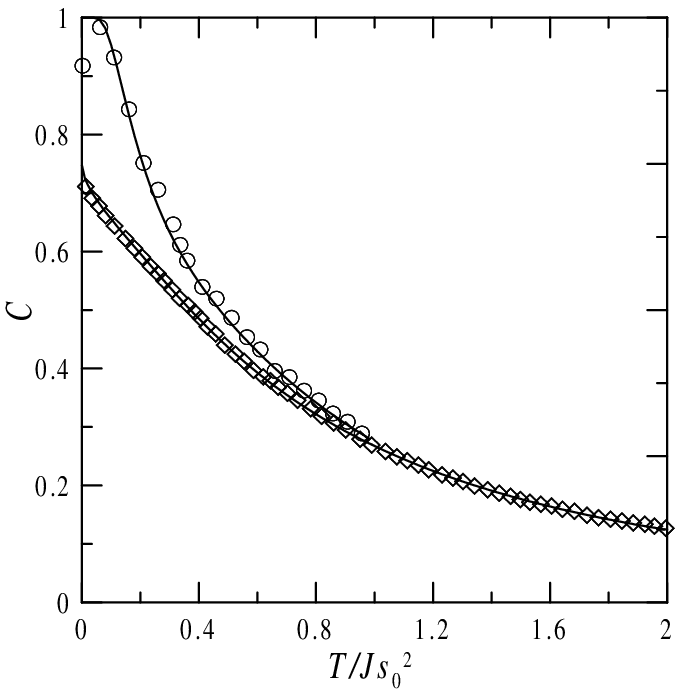} \par}
\caption{\label{fig.specific.heat}Specific heat from Monte Carlo simulations and the
one predicted by the GCC model: (open circles) Specific heat of the \emph{kagomé}
lattice \cite{Chalker.1992}; (open diamonds) Specific heat of the pyrochlore
lattice \cite{Reimers.1992}. The solid lines are the results of the GCC model.}
\end{figure}

In order to further check the accuracy of the model, we decided to compare its
predictions for the specific heat. The results are depicted in Fig. \ref{fig.specific.heat},
together with MC data \cite{Reimers.1992,Chalker.1992} for these systems. Again,
the agreement is excellent. For the pyrochlore lattice, the calculated specific heat is essentially exact down to 0 K, where a value of 3/4 is obtained \cite{moessner.1998}. However, for the \emph{kagomé} lattice, the exact value is \( 11/12 \), whereas our model predicts a value of 1. In any case, this deviation is expected to occur, as our model does not include the effect of zero modes, which are known to be especially important in the \emph{kagomé} case, where they give rise to the phenomenon known as order by disorder \cite{Chalker.1992,harris.1991}. In contrast, MC data suggest that this mechanism is absent in the Heisenberg pyrochlore.

\emph{Conclusions.-} In this work we have presented a generalization of the
so called constant coupling approach for geometrically frustrated antiferromagnets
with a pyrochlore or \emph{kagomé} structure. As a preliminary approach, we
have considered the most simple description of these systems, which consists
of a classical Heisenberg model with only nearest neighbors interactions. It
seems clear from Monte Carlo results that, in this approximation, these systems
remain paramagnetic in the whole temperature range. Analytical expressions for
the static magnetic susceptibility and specific heat can be obtained in the
framework of this model. These have been compared with Monte Carlo data for
these systems. The predicted susceptibility is essentially exact in whole the
temperature range for both types of lattice. The specific heat for the pyrochlore lattice is again found to be essentially exact down to zero temperature, whereas it deviates from the real value in the \emph{kagom\'{e}} case, as our model does not include the so called order by disorder phenomenon, which is especially important in this system at very low temperatures. It is specially remarkable that the curves
that are compared with Monte Carlo data do not contain any fit parameter, or
any rescaling factor.

Obviously, this is only a first step in the understanding of the uncommon features
present in frustrated geometries. There are still a lot of puzzling questions
regarding the nature of the spin glass and spin liquids observed in these systems,
which we do not think can be understood in the framework of the constant coupling
theory.

In any case, we think that the present model provides an excellent starting
point to understand open questions that remain even in the paramagnetic regime,
as are, to cite some examples, the effects of small anisotropies, dipole--dipole
interactions, or the effects of a small amount of dilution with non--magnetic
impurities in the magnetic lattice. Furthermore, the corresponding quantum generalized
constant coupling method provides some insight on how the quantum effects manifest themselves in the magnetic properties of these systems \cite{Garcia.2000}. It is important to note again, that the applicability of the present technique is not limited to the study of Heisenberg systems, but is useful for the investigation of the physical properties of other systems where geometrical frustration is relevant.
\begin{acknowledgments}

Angel García Adeva wants to thank M.A. Cazalilla for very stimulating discussions.
This work has been partially supported by the Spanish MEC, under the Subprograma
General de Formación de Personal Investigador en el Extranjero.

\end{acknowledgments}


\begin{thebibliography}{10}
\bibitem{observation}As commented in the introduction, the classical Heissenberg model with NN interactions
remains paramagnetic down to \( T=0 \) K. Therefore, a series expansion of
the magnetization is well justified in whole the temperature range, as long
as the applied field is small.
\end{thebibliography}
\end{document}